\documentclass[epj,nopacs]{svjour}
\usepackage{epsfig,tabularx,array,booktabs,calc,multirow,amsmath,color}
\usepackage{graphics}
\begin{document}
\title{Could $Z_{c}(4025)$ be a $J^{P}=1^{+}$ $D^{*}\bar{D^{*}}$ molecular state?}
\author{Chun-Yu Cui$^{\P}$, Yong-Lu Liu$^*$ and Ming-Qiu Huang$^*$}
\institute{$^{\P}$ Department of Physics, School of Biomedical Engineering, Third Military Medical University, Chongqing 400038, China\\
$^*$ College of Science, National University of Defense
Technology, Hunan 410073, China}
\date{\today}
\abstract{
We investigate whether the newly observed narrow resonance $Z_{c}(4025)$ can
be described as a $D^{*}\bar{D^{*}}$ molecular state with quantum numbers $J^{P}=1^{+}$.
Using QCD sum rules, we consider contributions up to dimension six in the operator product expansion
and work at leading order of $\alpha_{s}$. The mass obtained for this state is $(4.05\pm 0.28)~\mbox{GeV}$.
It is concluded that $D^{*}\bar{D^{*}}$ molecular state is a possible candidate for $Z_{c}(4025)$.
} 
\maketitle
%
\section{Introduction}\label{sec1}
The existence of charged states containing $c{\bar c}$ represents an indisputable signal of the exotic states.
There are three charged states $Z^+(4050)$, $Z^+(4250)$ and $Z^+(4430)$, which were discovered by the Belle collaboration in B-decays~\cite{BELLE,BELLE2}. This discovery has triggered many theoretical investigations on the nature of this structure~\cite{4430}. However, Babar did not confirm the existence of these three charged states~\cite{BABAR}. Recall that two charged bottomonium-like resonances $Z_{b}(10610)$ and $Z_{b}(10650)$ were observed by Belle Collaboration~\cite{10620}, most of theoretical investigations support the $B^{*}{\bar B^{(*)}}$ molecular structure with $J^{P}=1^{+}$~\cite{10610}. That implies there exist similar structures in charmonium-like energy regions. As anticipated, a charged charmonium-like structure, $Z_c(3900)$, was discovered by the BESIII experiment in the $\pi^\pm J/\psi$ mass spectrum in the process $e^+e^- \to \pi^+\pi^- J/\psi$~\cite{BEC}, and
subsequently confirmed by the Belle experiment~\cite{BELLE3}. Based on heavy quark spin symmetry and heavy flavour symmetry, Guo {\it et al.} predict $Z_{c}(3900)$ as the $D^{*}{\bar D}$ partner of the $Z_{b}(10610)$~\cite{Guo}.

Recently, the BECIII Collaboration reported a new enhancement structure $Z_c(4025)$ in the
$e^+e^-\to (D^*\bar{D}^*)^\pm\pi^\mp$ at $\sqrt{s}=4.26$ GeV~\cite{BEC2}.
The mass and width of this state is $M=(4026.3\pm2.6\pm3.7)~{\rm MeV}/c^2$ and $\Gamma=(24.8\pm5.6\pm7.7)~{\rm MeV}/c^2$. It attracts many attempts to investigate its possible configurations with various models~\cite{Wang,He,Chen,Qiao,KPK,Tor}. 
In Ref.~\cite{Wang}, they give an explanation of $Y(4260)\to (D^*\bar{D}^*)^-\pi^+$ decay via
the ISPE mechanism. In Ref.~\cite{He}, the authors have studied the loosely bound $D^{*}\bar{D^{*}}$ system with one-pion exchange model, which indicates that $Z_c(4025)$ is the ideal candidate of the $D^{*}\bar{D^{*}}$ molecular state with $J^{P}=1^{+}$. We also notice that, charmonium-like charged tetraquark state with $J^{P}=1^{+}$ was studied using QCD sum rules. The mass obtained is about $(4.05\pm0.05)$ GeV~\cite{Chen1}.

Due to the asymptotic property of the QCD, studies of the hadron physics have to concern about the nonperturbative effect which is difficult in quantum field theory. There are many methods to estimate the mass of a hadron, among which QCD sum rule(QCDSR)~\cite{svz,reinders,overview2,overview3,NielsenPR} is a fairly reliable one. In Ref.~\cite{Cui}, by assuming $Z_{c}(3900)$ as a $D\bar{D^{*}}$ molecular state with $J^{P}=1^{+}$, we investigate the mass of this possible molecular configuration within the framework of QCDSR. Although the $J^P$ quantum numbers of $Z_c(4025)$ remain to be determined, it is still preferred to have spin parity $J^{P}=1^{+}$ experimentally~\cite{BEC}. Following the opinion in Refs.~\cite{BEC,He}, we propose that $Z_c(4025)$ is a $D^{*}\bar{D^{*}}$ molecular candidate with $J^{P}=1^{+}$, which is the $Z_{b}(10650)$'s corresponding particle in the charmonium sector.
It is difficult to construct a suitable axial-vector style molecular current interpolating the state $Z_c(4025)$ using both $D^{*}$ and $\bar{D^{*}}$ fields. A possible interpolating field is supposed to describe this kind of state $D^{*}\bar{D^{*}}$:
\begin{eqnarray}\label{current}
j^{\mu}(x)&=&\varepsilon^{\mu\nu\alpha\beta}(\bar u(x)i\gamma_{\nu}c(x))D_{\alpha}(\bar{c
}(x)\gamma_{\beta}d(x)).
\end{eqnarray}
Performing the parity transformation to the current, it satisfies the condition $(j^{\mu})_p=j^{\mu}$.

The rest of the paper is organized as follows. The QCDSR for the $Z_{c}(4025)$ state is derived in Sec.~\ref{sec2}, with contributions up to dimension six in the operator product expansion(OPE). The numerical analysis is presented to extract the hadronic mass in Sec.~\ref{sec3}, where a short summary and conclusion are also presented.
\section{QCD sum rules for $Z_{c}(4025)$}\label{sec2}
In order to get the mass of the particle in the QCDSR approach, we start from considering the following two-point correlation function
\begin{eqnarray}
\Pi^{\mu\nu}(q^{2})=i\int
d^{4}x\mbox{e}^{iq.x}\langle0|T[j^{\mu}(x)j^{\nu+}(0)]|0\rangle.
\end{eqnarray}
In consideration of the Lorentz covariance, the above correlation function can be generally decomposed as
\begin{eqnarray}
\Pi^{\mu\nu}(q^{2})=(\frac{q^{\mu}q^{\nu}}{q^{2}}-g^{\mu\nu})\Pi^{(1)}(q^{2})+\frac{q^{\mu}q^{\nu}}{q^{2}}\Pi^{(0)}(q^{2}).
\end{eqnarray}
We select the term proportional to $g_{\mu\nu}$ to extract the mass sum rule, since it only gets contributions from the $1^{+}$ state. The QCD sum rule method attempts to link the hadron phenomenology with the interactions of quarks and gluons. Three main ingredients are contained in the process when using this method: an approximate description of the correlation function in terms
of intermediate states through the dispersion relation, an evaluation of the same correlation function in terms of QCD degrees of freedom via the operator product expansion (OPE), and a procedure for matching these two descriptions and extracting the parameters that characterize the hadronic state of interest.

We can insert a complete set of intermediate hadronic states with the same quantum numbers
as the current operators $j^{\mu}$ into the correlation function to obtain the
phenomenological expression. The coupling of the current with the state can be defined by the decay constant as follows:
\begin{eqnarray}
\langle 0|j_{\mu}|Z\rangle&=&\lambda \epsilon_{\mu}.
\end{eqnarray}
Therefore the invariant function $\Pi^{(1)}(q^{2})$ can be expressed as
\begin{eqnarray}\label{ph}
\Pi^{(1)}(q^{2})=\frac{\lambda^{2}}{M_{Z}^{2}-q^{2}}+\frac{1}{\pi}\int_{s_{0}}
^{\infty}ds\frac{\mbox{Im}\Pi^{(1)\mbox{phen}}(s)}{s-q^{2}},
\end{eqnarray}
where $M_{Z}$ denotes the mass of the molecular state, and $s_0$ is the threshold parameter.

In the OPE side, $\Pi^{(1)}(q^{2})$ can be written as
\begin{eqnarray}\label{ope}
\Pi^{(1)}(q^{2})=\int_{4m_{c}^{2}}^{\infty}ds\frac{\rho^{OPE}(s)}{s-q^{2}},
\end{eqnarray}
where the spectral density is $\rho^{OPE}(s)=\frac{1}{\pi}\mbox{Im}\Pi^{\mbox{(1)}}(s)$.
Applying the quark-hadron duality hypothesis with the Borel transformation, one obtains the following sum rule:
\begin{eqnarray}\label{sr}
\lambda^{2}e^{-M_{Z}^{2}/M^{2}}&=&\int_{4m_{c}^{2}}^{s_{0}}ds\rho^{OPE}(s)e^{-s/M^{2}},
\end{eqnarray}
with $M^2$ being the Borel parameter.

Technically, we work at the leading order of $\alpha_{s}$
and consider vacuum condensates up to dimension six, in use of the similar
techniques in Refs.~\cite{technique}. Considering the isospin breaking effect, we keep the terms which are linear in the light-quark masses $m_{u}$ and $m_{d}$. After some tedious calculations, the concrete forms of spectral densities can be derived:
\begin{eqnarray}
\rho^{OPE}(s)&=&\rho^{\mbox{pert}}(s)+\rho^{\langle\bar{q}q\rangle}(s)+\rho^{\langle
g^{2}G^{2}\rangle}(s)+\rho^{\langle
g\bar{q}\sigma\cdot G q\rangle}(s)\nonumber\\&&{}
+\rho^{\langle\bar{q}q\rangle^{2}}(s)+\rho^{\langle g^{3}G^{3}\rangle}(s),
\end{eqnarray}
with
\begin{eqnarray}\label{spectral}
\rho^{\mbox{pert}}(s)&=&\frac{m_{c}^{2}}{2^{12}\pi^{6}}\int_{\alpha_{min}}^{\alpha_{max}}\frac{d\alpha}{\alpha^{4}}\int_{\beta_{min}}^{1-\alpha}\frac{d\beta}{\beta^{3}}(1-{\alpha}-{\beta})^2
\nonumber\\&&{}\times(7{\alpha} +4{\beta}+8)r(m_{c},s)^{4}
\nonumber\\&&{}-\frac{1}{5*2^{12}\pi^{6}}\int_{\alpha_{min}}^{\alpha_{max}}\frac{d\alpha}{\alpha^{4}}\int_{\beta_{min}}^{1-\alpha}\frac{d\beta}{\beta^{4}}
\nonumber\\&&{}\times(29{\alpha}^3+60{\alpha}^2\beta+39{\alpha}\beta^2+8\beta^3-6{\alpha}^2
\nonumber\\&&{}-12{\alpha}\beta
-27\alpha-24\beta+16)r(m_{c},s)^{5}
,\nonumber\\
\rho^{\langle\bar{q}q\rangle}(s)&=&\frac{\langle\bar{q}q\rangle}{2^{6}\pi^{4}}m_{c}\int_{\alpha_{min}}^{\alpha_{max}}\frac{d\alpha}{\alpha(1-\alpha)^2}[m_{c}^{2}-\alpha(1-\alpha)s]^3
\nonumber\\&&{}+\frac{3\langle\bar{q}q\rangle}{2^{7}\pi^{4}}m_{c}^{3}\int_{\alpha_{min}}^{\alpha_{max}}\frac{d\alpha}{\alpha^{2}}\int_{\beta_{min}}^{1-\alpha}\frac{d\beta}{\beta^{2}}
\nonumber\\&&{}\times(\alpha^2+3\alpha\beta+2\beta^2-4\alpha-3\beta+1)r(m_{c},s)^{2}
\nonumber\\&&{}-\frac{\langle\bar{q}q\rangle}{2^{7}\pi^{4}}m_{c}\int_{\alpha_{min}}^{\alpha_{max}}\frac{d\alpha}{\alpha^{2}}\int_{\beta_{min}}^{1-\alpha}\frac{d\beta}{\beta^{3}}
\nonumber\\&&{}\times(4\alpha^2+8\alpha\beta+3\beta^2-10\alpha-4\beta+2)r(m_{c},s)^{3}
,\nonumber\\
\rho^{\langle g^{2}G^{2}\rangle}(s)&=&\frac{\langle
g^{2}G^{2}\rangle}{2^{12}\pi^{6}}m_{c}^{4}\int_{\alpha_{min}}^{\alpha_{max}}\frac{d\alpha}{\alpha^4}\int_{\beta_{min}}^{1-\alpha}d{\beta}
\nonumber\\&&{}\times(1-\alpha-\beta)^2(2\alpha+\beta+1)r(m_{c},s)
\nonumber\\&&{}-\frac{\langle
g^{2}G^{2}\rangle}{2^{13}\pi^{6}}m_{c}^{2}\int_{\alpha_{min}}^{\alpha_{max}}\frac{d\alpha}{\alpha^4}\int_{\beta_{min}}^{1-\alpha}\frac{d\beta}{\beta}
\nonumber\\&&{}\times(14{\alpha}^3+27{\alpha}^2\beta+16{\alpha}\beta^2+3\beta^3+14{\alpha}^2
\nonumber\\&&{}+4{\alpha}\beta-2\beta^2-20\alpha-5\beta+4)r(m_{c},s)^{2},\nonumber\\
\rho^{\langle g\bar{q}\sigma\cdot G q\rangle}(s)&=&\frac{3\langle
g\bar{q}\sigma\cdot G q\rangle}{2^{7}\pi^{4}}m_{c}\int_{\alpha_{min}}^{\alpha_{max}}d\alpha\alpha{s}[m_{c}^{2}-\alpha(1-\alpha)s]
\nonumber\\&&{}-\frac{3\langle
g\bar{q}\sigma\cdot G q\rangle}{2^{10}\pi^{4}}m_{c}\int_{\alpha_{min}}^{\alpha_{max}}\frac{d\alpha}{\alpha(1-\alpha)^2}
\nonumber\\&&{}\times(7\alpha^2-25\alpha+10)[m_{c}^{2}-\alpha(1-\alpha)s]^2
\nonumber\\&&{}
-\frac{3\langle g\bar{s}\sigma\cdot
Gs\rangle}{2^{8}\pi^{4}}m_{c}^3\int_{\alpha_{min}}^{\alpha_{max}}\frac{d\alpha}{\alpha}\int_{\beta_{min}}^{1-\alpha}\frac{d\beta}{\beta}
\nonumber\\&&{}\times(2-2\alpha-\beta)r(m_{c},s)
\nonumber\\&&{}
-\frac{3\langle g\bar{s}\sigma\cdot
Gs\rangle}{2^{9}\pi^{4}}m_{c}\int_{\alpha_{min}}^{\alpha_{max}}\frac{d\alpha}{\alpha^2}\int_{\beta_{min}}^{1-\alpha}\frac{d\beta}{\beta^2}
\nonumber\\&&{}\times(\alpha+\beta)r(m_{c},s)^2
,\nonumber\\
\rho^{\langle\bar{q}q\rangle^{2}}(s)&=&-\frac{\langle\bar{q}q\rangle^{2}}{2^{4}\pi^{2}}m_{c}^{2}\int_{\alpha_{min}}^{\alpha_{max}}d\alpha[m_{c}^{2}-\alpha(1-\alpha)s],\nonumber\\
\end{eqnarray}


\section{Numerical analysis and Summary}\label{sec3}
For numerical analysis of Eq. (\ref{sr}), we first specify the input parameters. The quark masses are chosen as $m_{u}=2.3~\mbox{MeV}$, $m_{d}=6.4~\mbox{MeV}$, and $m_{c}=(1.23\pm0.06)~\mbox{GeV}$ \cite{PDG}. The condensates are $\langle\bar{u}u\rangle=\langle\bar{d}d\rangle=\langle\bar{q}q\rangle=-(0.23\pm0.03)^{3}~\mbox{GeV}^{3}$,
$\langle g\bar{q}\sigma\cdot G
q\rangle=m_{0}^{2}~\langle\bar{q}q\rangle$,
$m_{0}^{2}=0.8~\mbox{GeV}^{2}$, $\langle
g^{2}G^{2}\rangle=0.88~\mbox{GeV}^{4}$, and $\langle
g^{3}G^{3}\rangle=0.045~\mbox{GeV}^{6}$~\cite{overview2}.
Complying with the standard procedure of the QCDSR, the threshold $s_{0}$ and Borel parameter $M^{2}$ are varied to obtain the optimal stability window. There are two criteria (pole dominance and convergence of the OPE) for choosing the Borel parameter $M^{2}$ and threshold $s_{0}$.

\begin{figure}
\centering
\includegraphics[totalheight=5cm,width=7cm]{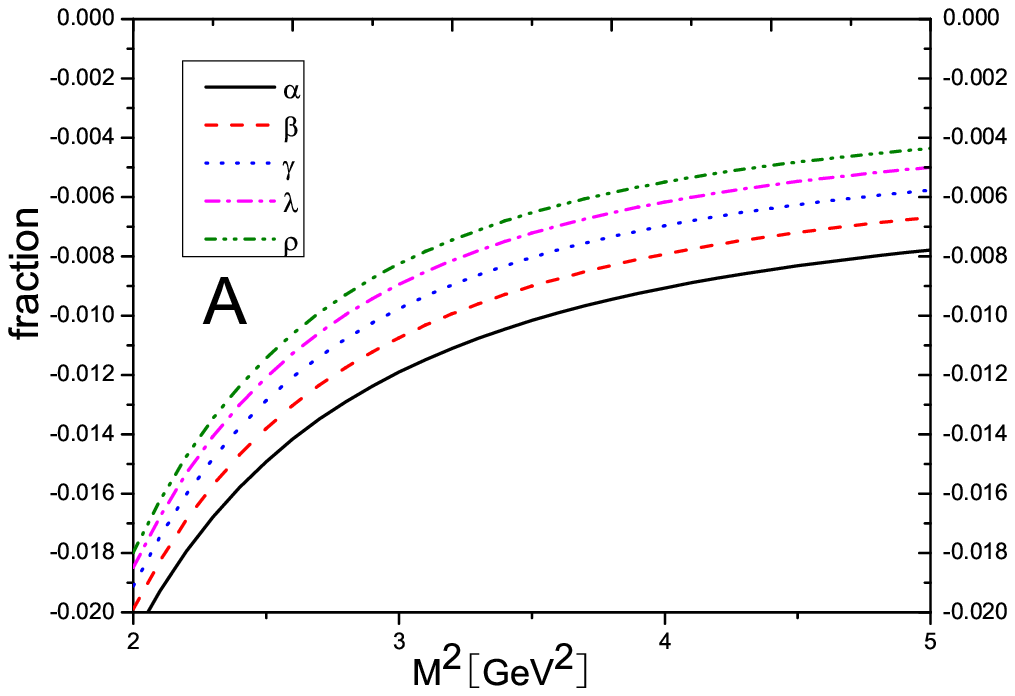}
\includegraphics[totalheight=5cm,width=7cm]{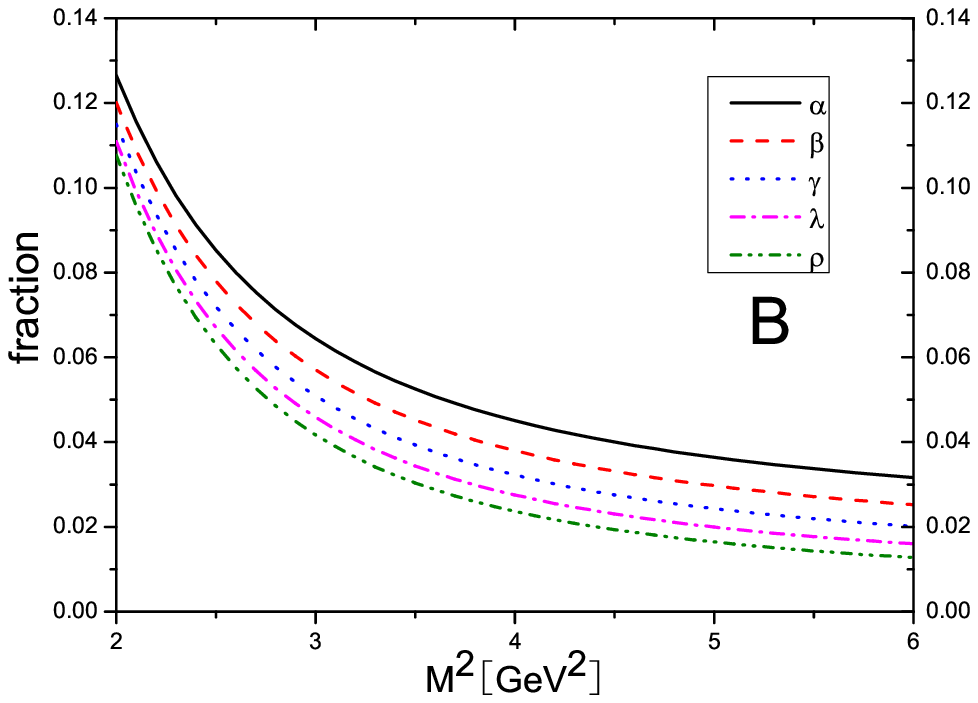}
\caption{The OPE convergence for the molecular state with contributions from different terms by varying the Borel parameter $M^2$. The $A$ and $B$ correspond to contributions from the $D=6$ term and the $D=5$ term, respectively.  The notations $\alpha$, $\beta$, $\gamma$, $\lambda$ and $\rho$  correspond to the threshold parameters $\sqrt{s_0}=4.5\,\rm{GeV}$, $4.6\,\rm{GeV}$, $4.7\,\rm{GeV}$, $4.8\,\rm{GeV}$ and $4.9\,\rm{GeV}$, respectively.}\label{fig1}
\end{figure}

\begin{figure}
\centerline{\epsfysize=6.0truecm\epsfbox{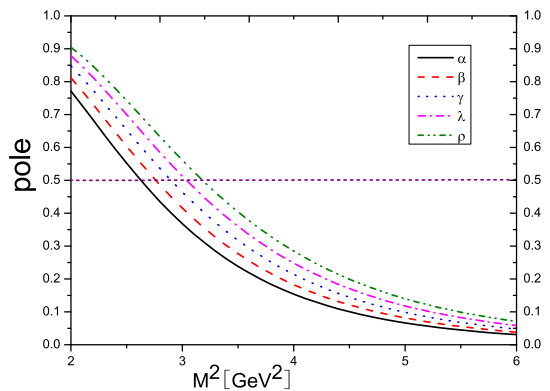}}
\caption{Contributions from the pole terms with variation of the Borel parameter $M^2$. The notations $\alpha$, $\beta$, $\gamma$, $\lambda$ and $\rho$  correspond to the threshold $\sqrt{s_0}=4.5\,\rm{GeV}$, $4.6\,\rm{GeV}$, $4.7\,\rm{GeV}$, $4.8\,\rm{GeV}$ and $4.9\,\rm{GeV}$, respectively.}\label{fig2}
\end{figure}

\begin{figure}
\centerline{\epsfysize=6.0truecm
\epsfbox{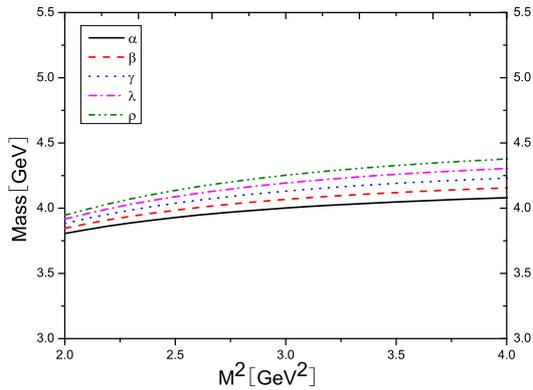}}\caption{
The mass of the molecular state as
a function of $M^2$ from sum rule (\ref{sr}). The notations $\alpha$, $\beta$, $\gamma$, $\lambda$ and $\rho$ correspond to the threshold parameters $\sqrt{s_0}=4.5\,\rm{GeV}$, $4.6\,\rm{GeV}$, $4.7\,\rm{GeV}$, $4.8\,\rm{GeV}$ and $4.9\,\rm{GeV}$, respectively.}\label{fig3}
\end{figure}

The contributions from the high dimension vacuum condensates in the OPE are shown in Fig.\ref{fig1}, as a function of $M^{2}$. We have used $\sqrt{s_0}\geq 4.5\,\mbox{GeV}$. From this figure it can be seen that for $M^2\geq 2.0\,\mbox{GeV}^2$, the contribution of the dimension-$6$ condensate is less than $3\%$ of the total contribution and  the contribution of the dimension-$5$ condensate is less than $13\%$ of the total contribution, which indicate the starting point for a good Borel convergence. Therefore, we fix the uniform lower value of $M^2$ in the sum rule window as $M^2_{min}= 2.0\,\mbox{GeV}^2$. The upper limit of $M^2$ is determined by imposing that the pole contribution should be larger than continuum contribution. In Fig.\ref{fig2}, we show the $M^{2}$ dependence of the contributions from the pole terms. Table~\ref{tab:1} shows the values of $M^2_{max}$ for several values of $\sqrt{s_0}$. In Fig.\ref{fig3}, we show the molecular state mass, for different values of $\sqrt{s_0}$, in the relevant sum rule window. It can be seen that the mass is stable in the Borel window with the corresponding threshold $\sqrt{s_0}$. The final estimate of the $J^{P}=1^{+}$ molecular state is obtained as
\begin{eqnarray}
M_{Z} = (4.05\pm 0.28)~\mbox{GeV}.
\label{Zmass}
\end{eqnarray}

\begin{table}
\caption{Upper limits in the Borel window for the $J^{P}=1^{+}$ $D^{*}\bar{D^{*}}$ current obtained from the sum rule for different values of $\sqrt{s_0}$.}\label{tab:1}
\begin{center}
\begin{tabular}{|c|c|}
\hline
$\sqrt{s_0}~(\mbox{GeV})$ & $M^2_{max}(\mbox{GeV}^2)$\\ \hline
4.5 &2.62 \\ \hline
4.6 &2.75 \\ \hline
4.7 &2.91 \\ \hline
4.8 &3.12 \\ \hline
4.9 &3.22 \\ \hline
\end{tabular}
\end{center}
\end{table}

In summary, we construct a possible interpolator to describe the $Z_{c}(4025)$ as a axial-vector $D^{*}\bar{D^{*}}$ molecular state. QCD sum rule approach has been applied to calculate the mass of the resonance.
Our numerical result is $M_{D^{*}\bar{D^{*}}} = (4.05\pm 0.28)~\mbox{GeV}$ which is compatible with the experimental data of $Z_{c}(4025)$ by BECIII Collaboration. Thus it is concluded that $Z_{c}(4025)$ may be a $D^{*}\bar{D^{*}}$ molecular state with quantum number $J^P=1^+$, which is supposed to be the charmonium-like partner of $Z_b(10650)$.

\section*{Acknowledgements}
This work was supported in part by the National Natural Science
Foundation of China under Contract Nos.11275268 and 11105222.

\end{document}